\documentclass[aps,prd,twocolumn,nofootinbib,floatfix]{revtex4-1}
\usepackage{setspace}
\usepackage{amsthm}
\theoremstyle{definition}

\usepackage{graphicx}
\usepackage{dcolumn}
\usepackage{multirow}
\usepackage{subfigure}
\usepackage{times,mathptm}
\usepackage{float}
\usepackage{color}
\usepackage{amsmath,amsfonts}
\usepackage{mathptmx}
\usepackage{mathrsfs}
\usepackage{bbm}
\usepackage{bm}
\usepackage{xfrac}

\newcommand{\beq}{\begin{equation}}
\newcommand{\eeq}{\end{equation}} 
\newcommand{\bea}{\begin{eqnarray}}
\newcommand{\eea}{\end{eqnarray}} 

\newcommand{\Om}{\Omega}
\newcommand{\up}{\uparrow}
\newcommand{\dn}{\downarrow}

\newcommand{\U}{{\cal U}}
\newcommand{\ua}{\uparrow}
\newcommand{\da}{\downarrow}

\newcommand{\E}{\mathcal{E}}

\renewcommand{\d}{\delta}

\renewcommand{\l}{\lambda}

\newcommand{\p}{\psi}
\renewcommand{\b}{\beta}
\renewcommand{\a}{\alpha}

\newcommand{\bx}{\mathbf{x}}

\newcommand{\bk}{\mathbf{k}}

\newcommand{\m}{\mu}

\newcommand{\g}{\gamma}
\renewcommand{\r}{\rho}
\newcommand{\e}{\epsilon}

\newcommand{\bp}{\mathbf p}

\newcommand{\oh}{\frac{1}{2}}

\newcommand{\dg}{\dagger}
\newcommand{\non}{\nonumber}

\newcommand{\rf}[1]{(\ref{#1})}
\newcommand{\ra}{\rightarrow}

\renewcommand{\vec}[1]{\bm #1}

\usepackage{ulem}

\begin{document}

\title{A Variational Improvement of the Hartree-Fock Approach to the 2D Hubbard Model}
\bigskip
\bigskip

\author{Kazue Matsuyama and Jeff Greensite}
\affiliation{Physics and Astronomy Department \\ San Francisco State
University  \\ San Francisco, CA~94132, USA}
\bigskip
\date{\today}
\vspace{60pt}
\begin{abstract}

\singlespacing

   We consider a refinement of the usual Hartree-Fock method applied to the 2D Hubbard model, in Nambu spinor formulation. 
The new element is the addition of a ``condensate inducing'' term proportional to a variational parameter $h$ to the 
Hartree-Fock Hamiltonian, which generates an s- or d-wave condensate at zero temperature.  This modified Hartree-Fock Hamiltonian is used only to generate variational trial states; energy expectation values are computed in the full two-dimensional Hubbard Hamiltonian with no modification.   It is found that there exist trial states  with non-vanishing condensates which are lower in energy than the standard Hartree-Fock ground states.   However, these lower energy condensate states exist only in a spatially inhomogeneous (stripe) phase.  No lowering of energy, relative to the Hartree-Fock ground state, is found in the spatially homogenous region of the $U-$density phase plane.

\end{abstract}

%
%
%
\maketitle
 
\singlespacing

\section{\label{Intro} Introduction}

   There is still some question about whether, and for which parameter values, the ground state of the two-dimensional 
Hubbard model exhibits the d-wave condensation observed in cuprate superconductors.  The model is difficult to 
solve convincingly due to the sign problem, and all existing methods seem to have their limitations, both theoretical and practical.  
The earliest approach,
dating to 1966, is the Hartree-Fock method applied to the Hubbard model, and this venerable approach has a very large literature,
a sample of which is  \cite{Penn,*Hirsch,*Poilblanc,*Zaanen,*Machida,*Schulz1,*Schulz2,*Ichimura,*Verges,*Inui,*Dasgupta,*Xu}. Two contributions of ours along these lines are in \cite{Matsuyama:2022kam,Matsuyama:2022dtd}.  The strengths and limitations of the Hartree-Fock method are well known, and are described in a number of reviews \cite{Scalettar,Powell,Lechermann,Imada,Fazekas}.    Like any mean field approach the method neglects correlations, and in addition uses a single Slater determinant to approximate what is surely a more
complex ground state.  Nevertheless, this method was successful in predicting stripe patterns (Zaanen and Gunnarsson, Poilblanc and Rice in \cite{Zaanen}) later observed in experiment \cite{Emery}, as well as the emergence of ferromagnetism and the relation to the Stoner criterion.  So it is of interest to see how far one can go with a single Slater determinant, in particular whether the existing Hartree-Fock approach can be significantly improved, and whether an improved approach can tell us anything about d-wave condensation and superconductivity.

   In this article we consider going a little beyond the standard Hartree-Fock approach by incorporating a variational element.  The idea is to modify the Hartree-Fock Hamiltonian  by adding a term proportional to a variational parameter $h$ which induces a d- or s-wave condensate, and then regarding the ground state of the modified Hartree-Fock Hamiltonian as a trial state for the unmodified Hubbard Hamiltonian.  The energy expectation values of the unmodified Hubbard Hamiltonian, at each $h$, are evaluated in the trial states, and it will be seen that
in some regions of the $U$-density plane the trial states, with a non-vanishing condensate, are lower in energy than the
standard Hartree-Fock ground state.  It is found that spatial inhomogeneity in spin expectation values (``stripes'' and other patterns)
is correlated with the lowering of energy by d- or s-wave condensates. No condensation is found in the spatially homogeneous
phase, at least in this variational approach.

\section{Hartree-Fock in Nambu spinor formalism}
   We begin with a review of some formalities.  The 2D Hubbard model in Nambu spinor formalism is \cite{Matsuyama:2022dtd},
\bea
 H   &=& -t \sum_{<xy>} \{\p^\dg_1(x) \p_1(y) - \p^\dg_2(x) \p_2(y)\}   \non \\
    & &+ U \sum_x \p^\dg_1(x) \p_1(x) 
       -U \sum_x \p^\dg_1(x) \p_1(x) \p^\dg_2(x) \p_2(x) \non \\
     & &   - \m  \sum_{x} \{\p^\dg_1(x) \p_1(x) + 1 - \p^\dg_2(x) \p_2(x)\} \ ,
\label{H}
 \eea
 where $\psi_1,\psi_2$ are the upper and lower components, respectively, of the Nambu spinor $\psi$, $\m$ is the chemical potential, and we neglect any hopping terms beyond nearest neighbor.  We recall the definitions
\bea
 \psi(x) &=& \left[ \begin{array}{c} c_\ua(x) \cr  \cr  c^\dg_\da(x) \end{array} \right] 
    ~~~,~~~    \psi^\dg(x) = [c^\dg_\ua(x) , c_\da(x) ] \ ,\non \\
 \eea
where $c^\dg_s(x),c_s(x)$ are the usual electron creation/destruction operators.
 The trial (``Slater determinant'') state is
\beq
         |\Om \rangle = \prod_{i=1}^M \bigg( \sum_{x_i,\a_i} \phi_i(x_i,\a_i) \psi^\dg_{\a_i}(x_i) \bigg) |0\rangle \ .
 \label{Om}
 \eeq
 Note that the Fock vacuum $|0\rangle$ in Nambu formulation, annihilated by all $\psi$ operators, differs from the no-electron Fock vacuum in standard formulation.
 $|\Om\rangle$ is not an eigenstate of particle number, in the Nambu formulation, but it is an eigenstate of the
difference of electron spin up/down number operators $N_\up-N_\dn$. Setting $M=L^2$ in \rf{Om} restricts to states with an absence of 
overall magnetic moment (equal numbers of spin up and down), with the expectation value of
electron density controlled by the chemical potential $\m$.
In the Hartree-Fock approach the one-particle wave functions  $ \phi_i(x_i,\a_i)$ are to be determined self-consistently, and defining
 \bea
 \r(x,\a\b) &=& \langle \Om| \psi^\dg_\a(x) \psi_\b(x) |\Om\rangle \non \\
               &=& \sum_i \phi^*_i(x,\a) \phi_i(x,\b) \ ,
\label{rho}
\eea
the Hartree-Fock Hamiltonian is
\begin{widetext}
\bea
 H_{HF}(\Om)
  &=& -t \sum_{<xy>} ( \psi_1^\dg(x) \psi_1(y) - \psi_2^\dg(x) \psi_2(y)) 
   - \m \sum_{x} ( \psi_1^\dg(x) \psi_1(x) +1 - \psi_2^\dg(x) \psi_2(x)) \non \\
  & & + U \sum_x  \bigg\{ [1-\r(x,22)] \psi^\dg_1(x) \psi_1(x)  - \r(x,11) \psi^\dg_2(x) \psi_2(x) 
  + \r(x,12) \psi^\dg_2 \psi_1 + \r(x,21)\psi_1^\dg \psi_2 \bigg\} \non \\
  & & + U\sum_x \{\rho(x,11) \rho(x,22) - \rho(x,12) \rho(x,21) \} \non \\
  &=& \sum_x \sum_y \psi^\dg_\a(x) [H]_{x\a,y\b} \psi_\b(y) \ .
\label{Heff3}
\eea
\end{widetext}
Then the one-particle wavefunctions in the Slater determinant satisfy
\beq
 [H]_{x\a,y\b}  \phi_i(y,\b) =  E_i \phi_i(x,\a) \ .
\eeq
When the one particle energies $E_i$ are arranged, for convenience, in order of increasing energy, then the Slater determinant
state \rf{Om} with $M=L^2$ is the zero temperature ground state of the Hubbard model in the Hartree-Fock approximation.   The pairing condensate in momentum space is
\bea
        P(\bp) &=& {1\over L^2} \sum_x \sum_{x'} \langle \Om | c^\dg_\ua(x) c^\dg_\da(x') |\Om \rangle e^{i \bp \cdot(\bx-\bx')} \non \\
        &=&  {1\over L^2} \sum_x \sum_{x'} \langle \Om | \psi^\dg_1(x) \psi_2(x') |\Om \rangle e^{i \bp \cdot(\bx-\bx')} \non \\ 
        &=& {1\over L^2} \sum_x \sum_{x'}  \sum_{i=1}^{L^2} \phi_i^*(x,1) \phi_i(x',2) e^{i \bp \cdot(\bx-\bx')} 
         \non \\
         &=&  {1\over L^2}  \sum_{i=1}^{L^2} \phi_i^*(p,1) \phi_i(p,2) 
\label{pc}
\eea
on an $L\times L$ lattice volume.

There is no obvious evidence of a condensate in the ground state of the Hartree-Fock Hamiltonian, in either conventional \cite{Penn,*Poilblanc,*Zaanen,*Machida,*Schulz1,*Schulz2,*Ichimura,*Verges,*Inui,*Dasgupta,*Xu}
 or Nambu spinor \cite{Matsuyama:2022dtd} formulation.  In general, in Nambu formulation, a condensate exists if there
 are one-particle states in the Slater determinant such that $\phi_i(x,\a) \ne 0$ for both indices $\a=1,2$, as we see in eq.\ \rf{pc}.  But it is clear from inspection of \rf{Om}, \rf{rho}, \rf{Heff3}  that Hartree-Fock self-consistency allows that each $\phi_i(x,\a)$ is nonzero for only one of the two indices
 $\a=1,2$ (and $\rho(x,12)=\rho^*(x,21)=0$).  In that case there is no condensate in this mean field approach.
 
 \section{Variational approach: spatial invariance}
 With this in mind, let us modify $H_{HF}$ by adding a ``condensate inducing'' term to the matrix $[H]_{x\a,x'\b}$
 \bea
        [V]^{cond}_{x\a,x'\b} &=&   h(\d_{\a 1}\d_{\b 2}+\d_{\a 2}\d_{\b 1}) \non \\
        &\times& (\d_{x,x'+\hat{e}_x} + \d_{x,x'-\hat{e}_x} + q\d_{x,x'+\hat{e}_y} + q\d_{x,x'-\hat{e}_y}) \ ,
 \eea
 where $\hat{e}_x, \hat{e}_y$ are unit vectors in the $x,y$ directions respectively, $h$ is a
 variational parameter, and $q=\pm 1$, where $q=+1$ induces an s-wave condensate, and $q=-1$ induces a d-wave condensate.  We will denote the modified Hartree-Fock Hamiltonian as $H_{HFq}$.  Let us regard the ground state of $H_{HFq}$, denoted  $|\Om_h\rangle$, as a trial wave functional, dependent on the variational parameter $h$. If  the expectation value of energy density 
 \beq
 \E(h)={1\over L^2} \langle \Om_h|H|\Om_h\rangle
 \eeq 
 of the full Hubbard Hamiltonian \rf{H} is minimized at some ${h>0}$, then by the rules of the variational approach the ground state $|\Om_h\rangle$ is a better approximation, as compared to the Hartree-Fock ground state $|\Om_0\rangle$, to the true ground state of the Hubbard Hamiltonian.  Then the procedure is to choose some set of
parameters $\{U,\m\}$ (setting $t=1$ for numerical work), and at each $U,\m$ and variational parameter $h$ we compare $\E(0)$ to (i) $\E(h)$ for the ground state $|\Om_h\rangle$ of $H_{HF-}$, associated with a d-wave condensate; and (ii) $\E(h)$ for the ground state $|\Om_h\rangle$ of $H_{HF+}$, associated with an s-wave condensate.  If $h>0$ is energetically favorable, in cases (i) and/or (ii), we choose the value of $h$ and sign of $q$ which gives the lowest $\E$.  Whichever ground state has the lowest $\E$ is the state which is favored at the given $U,\m$, at the electron density
\beq
       f = {1\over L^2} \sum_x (\r(x,11) + 1 - \rho(x,22)) \ .
\label{f}
\eeq
We emphasize that the $h$ parameter is absent in the Hubbard Hamiltonian $H$, and only concerns the choice of trial ground state
$|\Om_h\rangle$.

 We will say that the system is spatially homogeneous if the $\r(x,\a\b)=\rho(\a\b)$ are position independent, which, within the
 Hartree-Fock approximation, is known to be true in lower density regions of the phase diagram.  In this case the Hamiltonian
 can be diagonalized in momentum space, and the calculation is greatly simplified.  In the absence of ferromagnetism there
 are equal numbers of spin up ($N_\up$) and spin down ($N_\dn$) electrons, and therefore
 we have, in Nambu formalism,
 \beq
 \rho(11)=1-\rho(22) \ ,
 \eeq
 with electron density
 \beq
  f=\rho(11)+1-\rho(22) =2\rho(11) \ .
 \eeq
 It is also consistent to assume $\rho(12)=\rho(21)=0$. 
 Translation invariance of $\rho$ implies that $[H]$ is diagonalized by momentum eigenstates, and on an 
$L\times L$ lattice, in the basis
\beq
 \phi_1(k) = {1\over L} \left[ \begin{array}{c} e^{i\vec{k}\cdot \vec{x}} \cr 0 \end{array} \right] ~~~,~~~
 \phi_2(k) = {1\over L} \left[ \begin{array}{c} 0 \cr e^{i\vec{k}\cdot \vec{x}} \end{array} \right] \ ,
 \label{phi}
 \eeq
 the matrix $[H]$ is decomposed, at each $k$ and for $q=-1$, into blocks
 \bea
 [H]_k &=& \left[ \begin{array}{cc}
                     \e^1_k & 2 h (\cos(k_x)-\cos(k_y)) \cr
                      2 h (\cos(k_x)-\cos(k_y)) & \e^2_k \end{array} \right] \non \\ \non \\
            & & +   (U \rho(11)\rho(22) -\m) \mathbbm{1} \ ,
 \eea
 where
 \bea
 \e^1_k &=& -2t( (\cos(k_x)+\cos(k_y)) + U n_\da - \m \non \\
 \e^2_k &=&  \ \ \ 2t( (\cos(k_x)+\cos(k_y))  - U n_\ua + \m \ ,
 \eea
and where $n_\ua=\r(11);  n_\da=1-\r(22)$ are the expectation values of the number densities of up and down electrons, respectively.  In the absence of net magnetization, these numbers are equal.  In that case  $ \e^2_k = -  \e^1_k$.

   The matrix $[H_k]$ is readily diagonalized.  Denoting 
\bea
\e_k&=&\e^1_k=-2t( (\cos(k_x)+\cos(k_y)) + U n_\da - \m \non \\
\b_k&=&2 h (\cos(k_x)-q\cos(k_y)) \ ,
\eea
the eigenvalues are 
\beq
       \l_{k\pm} = \pm \sqrt{\e_k^2 + \b_k^2} + U \rho(11)\rho(22) -\m \ ,
\eeq
and the eigenstate of lower energy is
\bea
\varphi(k) &=&  \left[ \begin{array}{c} \varphi(k,1) \cr  \varphi(k,2) \end{array} \right] \non \\
                 &=&  {  \left[ \begin{array}{c} \e_k - \sqrt{\e_k^2 + \b_k^2} \cr  \b_k \end{array} \right]  \over 
                   \sqrt{\b_k^2 + \left(-\e_k +\sqrt{\e_k^2+\b_k^2}\right)^2} } \ .
\label{evectors}
\eea
 The condensate \rf{pc} is
 \beq
          P(k) = \varphi(k,1) \varphi(k,2) \ .
 \eeq

As $h$ and $\b_k\ra 0$ the eigenvectors at $\e_k \ne 0$ are simply
\bea
\varphi(k) &=&  \left[ \begin{array}{c} 0 \cr  1  \end{array} \right] , ~~~  \e_k > 0 \non \\
                &=&  \left[ \begin{array}{c} 1 \cr  0   \end{array} \right] , ~~~  \e_k < 0 \ ,
\eea
while at $\e_k=0$ we have
 \bea
 \varphi(k)       &=&  {1\over \sqrt{2}} \left[ \begin{array}{c} -1\cr 1  \end{array} \right] , ~~~  \e_k =0,~ \b_k/h>0 \non \\
                &=&   {1\over \sqrt{2}} \left[ \begin{array}{c} 1 \cr  1   \end{array} \right] , ~~~  \e_k =0, ~ \b_k/h<0  \ ,
\eea
and in this limit
\bea
P(\bk) =  \left\{ \begin{array}{cl} 
       \oh \text{sign}(\cos k_x - \cos k_y) & \bk \in \text{Fermi surface} \cr
       0 & \text{otherwise} \end{array} \right. \ .
\label{dwave}
\eea
We therefore seem to have a condensate even in the $h\ra 0$ limit.  This is due to an energy degeneracy on the Fermi surface ($\e_k=0$)
which is lifted by even an infinitesimal perturbation.
Since this seems to be a case where the slightest perturbation
breaks a ground state degeneracy, even as the strength of the breaking term goes to zero, it is tempting to regard this
situation as an example of spontaneous symmetry breaking.
In fact this view was already presented in \cite{Matsuyama:2022dtd}, but there are two objections, both connected with lattice volume $L^2$.  The first is that, unlike spontaneous symmetry breaking, this breaking exists already at finite volume; no special order of limits $L^2 \ra \infty$ followed by $h\ra 0$ is necessary.  But the more serious issue                                                                                                                                                                                                                                                                                                                                                                                                                                                                                                                                                                           
is that in the $h\ra 0$ limit the condensate only exists {\it exactly} at the Fermi surface. In that case the condensate density is proportional
to $1/L$, and vanishes in the infinite volume limit.

\begin{figure}[htb]
 \center
 \includegraphics[scale=0.4]{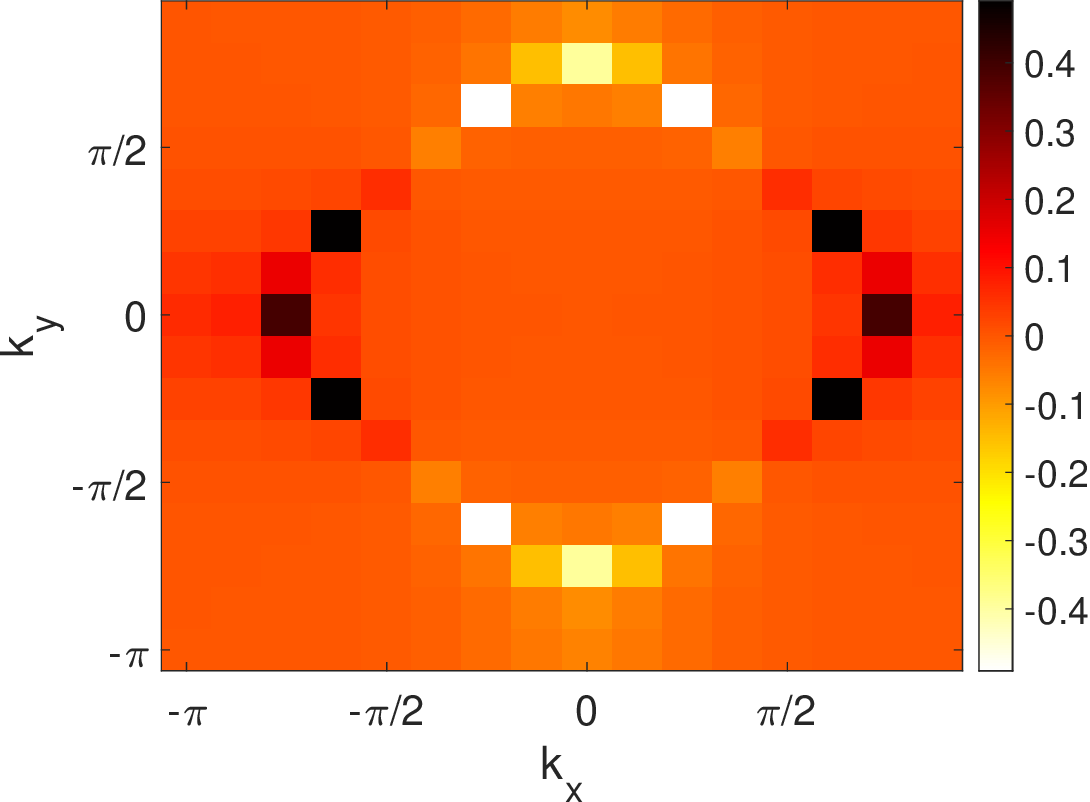}
 \caption{Momentum space condensate $P(k)$ at $U=2,\m=0.1$ and $h=0.002$, where we find density $f=0.72$. In this case,
 however, a lower energy is obtained at $h=0$.} 
 \label{pairU2}
 \end{figure}

    But there is no reason to take the $h\ra 0$ limit. The real objective is to find the
best possible approximation, within the limitations dictated by a single Slater determinant, to the ground state of the full Hubbard Hamiltonian of \rf{H}.  The question is whether eigenstates of the modified Hartree-Fock Hamiltonian $H_{HFq}$ would lower the expectation value of the Hubbard Hamiltonian $H$ at some $h>0$.  If so, then since one-particle wavefunctions would have non-zero amplitudes for both $\a$ index values, a condensate is obtained.  At finite $h$ the condensate density does
not vanish in the infinite volume limit, but is still concentrated in a small region of momentum space in the neighborhood of the
Fermi surface.  A typical result on a $16^2$ lattice is shown in Fig.\ \ref{pairU2}.

      But the question is whether the energy expectation value of the Hubbard Hamiltonian $\E(h)$ is less than $\E(0)$, for some
$h\ne 0$, where the energy densities are
\bea
         \E(h) &=& {1\over L^2} \langle \Om_h|H|\Om_h  \rangle   \non \\
                  &=& \sum_k \varphi^\dg(k)  \left[ \begin{array}{cc}
                     \e^1_k & 0\cr
                     0 & \e^2_k \end{array} \right] \varphi(k) + U \rho(11)\rho(22) -\m \ , \non \\ 
\eea
and the $\varphi(k)$ are given in \rf{evectors}.
Unfortunately this is not the case.  No matter the choice of $U,\m,h$, we always find ${\E(0)<\E(h)}$.  Condensate states
in the spatially homogenous region ($\rho(x,\a\b)$ $x$-independent) are not energetically preferred, at least in this extension
of the Hartree-Fock procedure.

\section{Variational approach: spatial inhomogeneity}

    Although it is true that regions of low density in the $U$-density plane are spatially homogenous in the sense defined above, it is
also well known that this is not generally the case at all densities, where stripes and other types of spatial inhomogenities
in $\rho(x,\a\b)$ are manifest.  So the question is whether such inhomogeneities can change the disappointing conclusion of the previous section.  To find out, we drop the assumption of spatial homogeneity and determine the ground state $|\Om_h\rangle$ of the modified Hartree-Fock Hamiltonian $H_{HFq}$ self-consistently, by standard numerical methods.\footnote{A useful check is to compare the results obtained by the methods of the previous section with the numerical approach that makes no assumption of spatially homogeneity, for
values of $U,\m$ where the latter method finds $\rho(x,\a\b)$ to be constant.  Since the answers must (and in fact do) agree in that case, this provides a modest check of our numerical code.}  Now
\beq
         |\Om_h \rangle = \prod_{i=1}^M \bigg( \sum_{x_i,\a_i} \phi^h_i(x_i,\a_i) \psi^\dg_{\a_i}(x_i) \bigg) |0\rangle \ .
 \label{Omh}
 \eeq
 where
 \beq
 \left([H]_{x\a,y\b}+[V^{cond}]_{x\a,y\b} \right)\phi^h_i(y,\b) =  E^h_i \phi^h_i(x,\a) \ .
\eeq
with
 \bea
 \r(x,\a\b) &=& \langle \Om_h| \psi^\dg_\a(x) \psi_\b(x) |\Om_h\rangle \non \\
               &=& \sum_i \phi^{h*}_i(x,\a) \phi^h_i(x,\b) \ ,
\label{rhoh}
\eea
determined self-consistently.  This defines the trial state.  

Defining also
\bea
\r_{nn}(x,\a\b) &=&  \sum_{i=1}^{L^2}  \phi^{*h}_i(x,\a)\bigg( \phi^h_i(x+\hat{e}_x,\b) + \phi^h_i(x-\hat{e}_x,\b) \non \\
   & & +\phi^h_i(x+\hat{e}_y,\b) + \phi^h_i(x-\hat{e}_y,\b) \bigg) \ ,
\eea
 we have
 \bea
 & & \lefteqn{\langle \Om_h |H| \Om_h \rangle = } \non \\
 & & ~~~ -t \sum_x \{ \rho_{nn}(x,11) - \rho_{nn}(x,22) \} + U \sum_x \r(x,11)   \non \\
 & & ~~~     - U\sum_x \bigg( \r(x,11) \r(x,22) -
                  \r(x,12) \r(x,21) \bigg) \non \\
  & & ~~~ - \m\sum_x \{ \rho(x,11) + 1 - \rho(x,22) \}  \ ,
 \label{EN}
 \eea
 and energy density
 \beq
      \E(h) =  {1\over L^2} \langle \Om_h| H |\Om_h\rangle \ .
\eeq
 
\begin{figure}[h!]
 \center
 \includegraphics[scale=0.7]{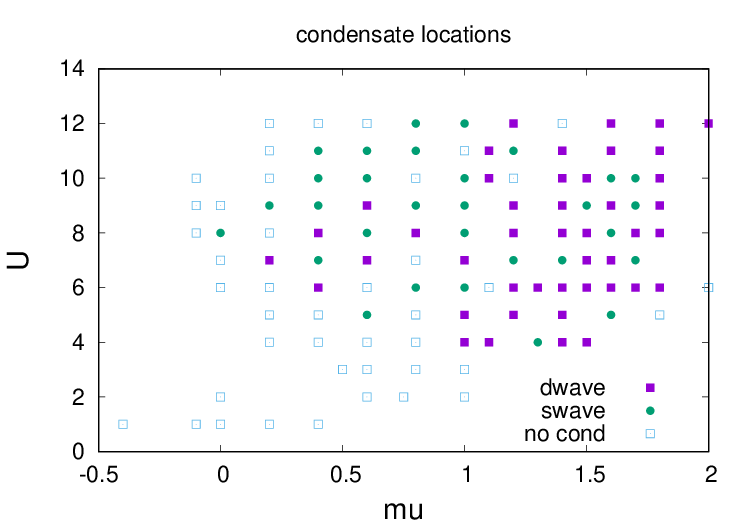}
 \caption{Numerical calculations were carried out the points indicated in the $U-\m$ plane.  Filled squares indicate d-wave condensates, filled circles indicate s-wave condensates, open symbols indicate that the no-condensate state at $h=0$ has the lowest energy expectation value $\E$.}
 \label{cond2}
 \end{figure}

\begin{figure}[h!]
 \center
 \includegraphics[scale=0.7]{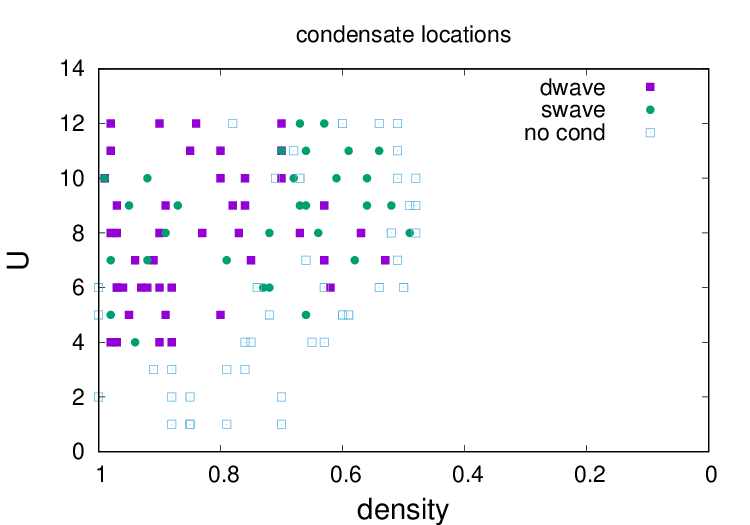}
 \caption{Data obtained at the $U,\m$ values shown in Fig.\ \ref{cond2} are each associated with a density $f$.  In this figure
the same points are plotted in the $U-f$ plane.}
 \label{cond1}
 \end{figure}

    The calculations were carried out on $16\times 16$ lattices with periodic boundary conditions, and our results are shown in the $U-\m$ plane in Fig.\ \ref{cond2}, and in the $U-f$ (coupling-density)  plane in Fig.\ \ref{cond1}, 
where $f=1$ is half-filling. Each
 of the symbols in Fig.\ \ref{cond2} represents a pair of $U,\m$ values where we have carried out the calculation described above.  At the filled square locations, $\E(h)$ for ${q=-1}$,
i.e.\ d-wave, is lower in energy than either $q=1$ (s-wave) or $h=0$ (standard Hartree-Fock).  At filled circle locations, $\E(h)$ is lowest for an s-wave condensate.  The open square symbols represent locations where standard Hartree-Fock, i.e.\ $h=0$ with no condensate, have the lowest energy.   The optimum $h$ varies, but is generally $O(10^{-2})$.  
Qualitatively, from Fig.\ \ref{cond1}, we see that the condensate disappears completely at low $U<4$, and at densities below half-filling.  Obviously, with a finite set of $\m$ values we have not covered the entire $U-\m$ plane, although nearby filled symbols at a given $U$ probably indicate a continuum of densities with condensate states.  We have carried out the calculation up to $U=12$.

 \begin{figure}[h!]
 \center
 \includegraphics[scale=0.5]{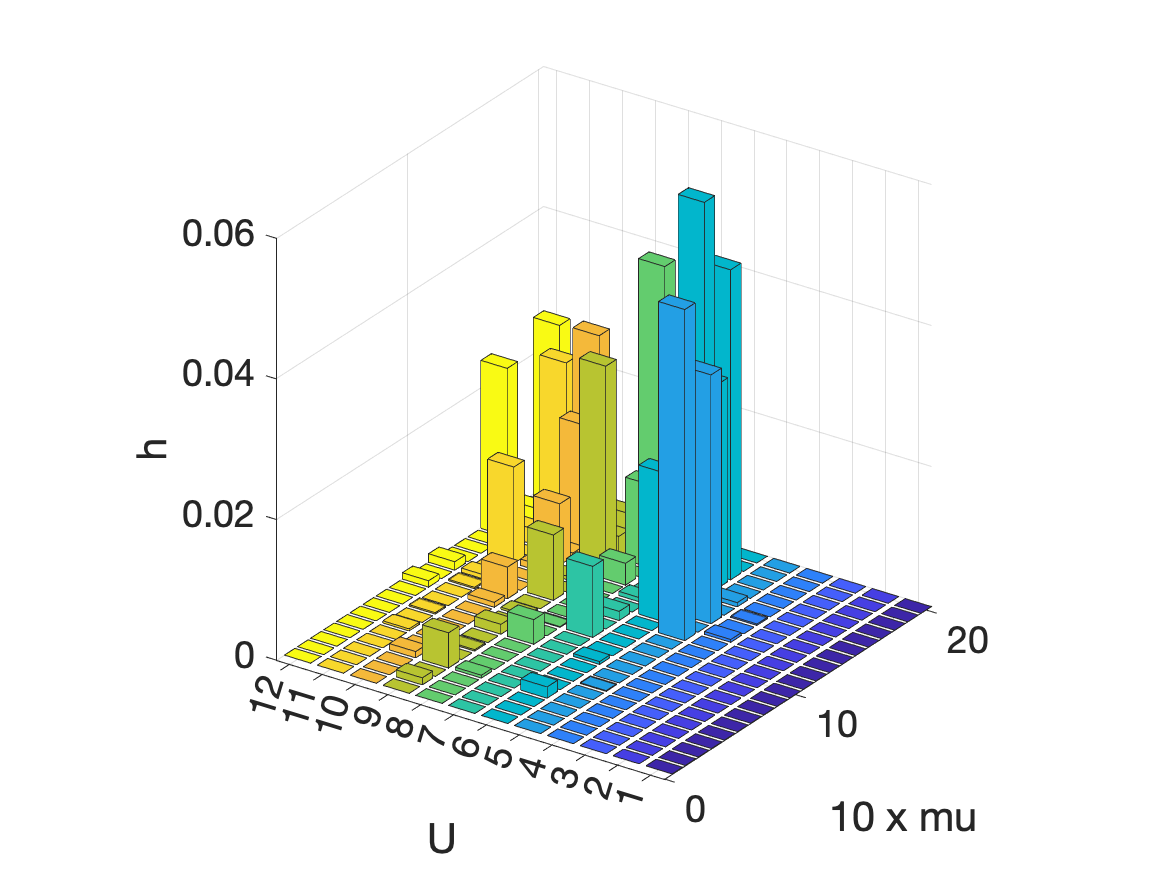}
 \caption{Optimal values of the variational parameter $h$ in the $U-\m$ plane for the state of lowest energy, at points where
 we find either s- or d-wave
 condensates. }
 \label{h}
 \end{figure}
 
\begin{figure}[h!]
 \center
 \includegraphics[scale=0.5]{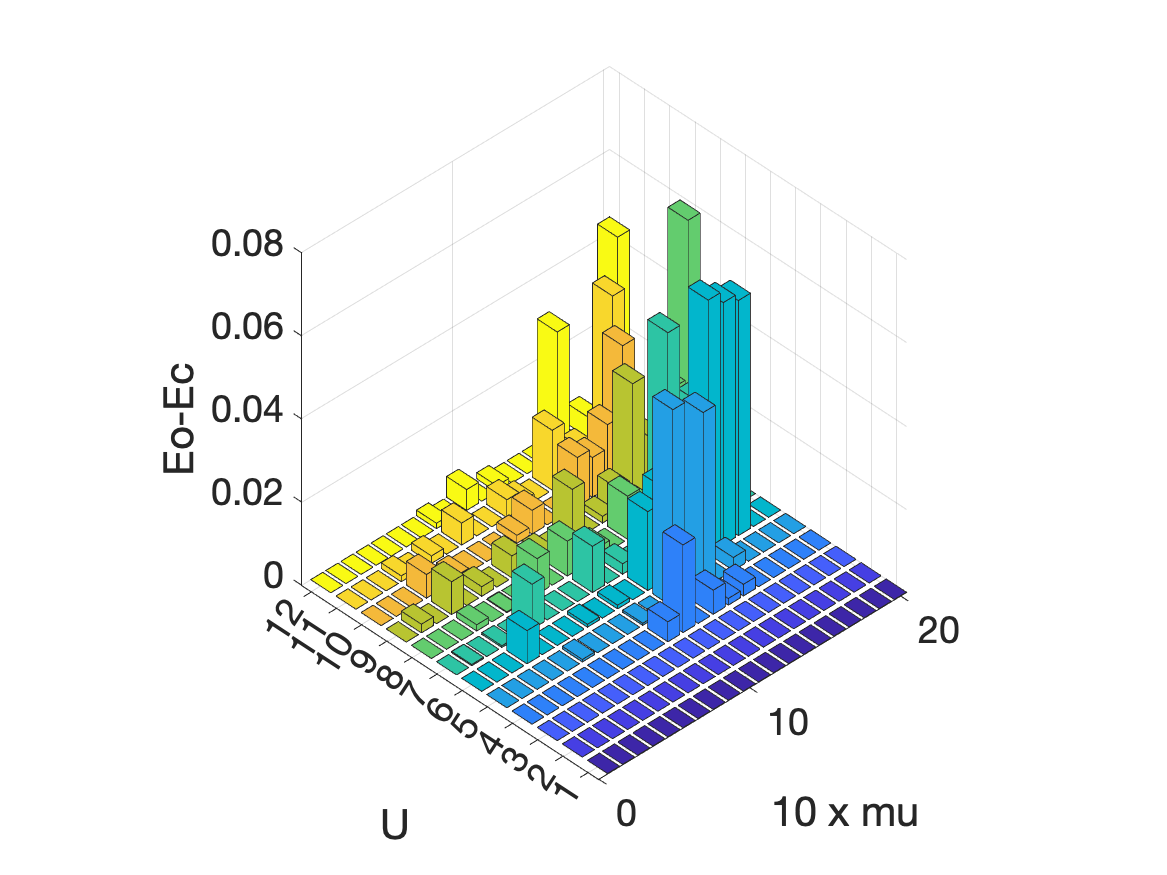}
 \caption{The difference $E_0-E_c$ between the $h=0$ energy $E_0$, and the energy $E_c$ of the optimal condensate ground state, at points
 where we find $E_c<E_0$.}
 \label{E0}
 \end{figure}

\begin{figure}[h!]
 \center
 \includegraphics[scale=0.5]{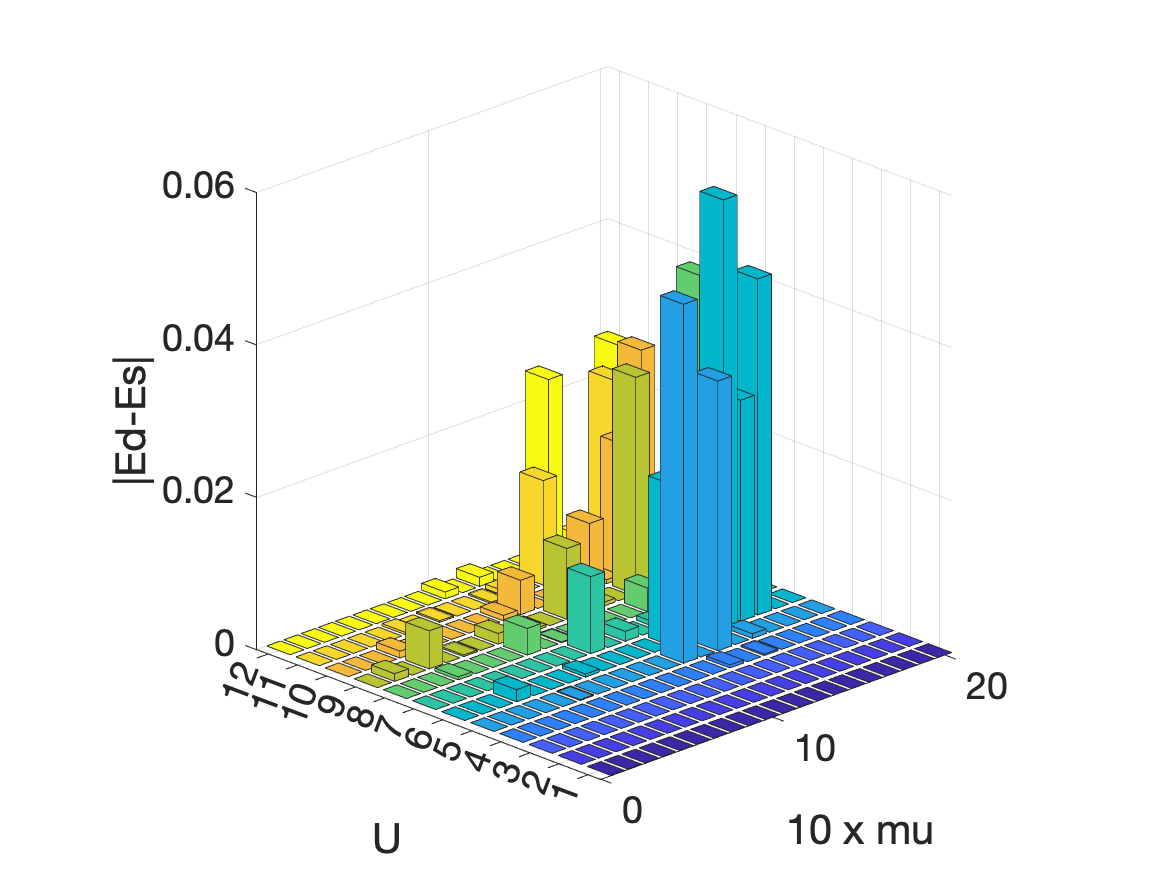}
 \caption{The magnitude of the difference between the optimal s- and d-wave ground state energies, when those energies are both
 less than the corresponding $E_0$.}  
 \label{dsdiff}
 \end{figure}

 The optimal values of $h>0$ in the $U-\mu$ plane, for the $U,\mu$ points which have been studied (see Fig.\ \ref{cond2}) 
 are shown in Fig.\ \ref{h}. Zero values, in this and in Figs.\ \ref{E0} and \ref{dsdiff}, are either points in which the unmodified Hartree-Fock state has the lowest energy, or else were not investigated.  Coordinates on the $x,y$ axes of these bar plots are required, by the plotting software, to be integer valued, and for this reason we have multiplied our $\mu$ values in this and the following two plots by a factor of ten.  Figure \ref{E0} displays the energy difference between the energy $E_0$ of the $h=0$ unmodified state, and the energy $E_c$ of the d- or s-wave state, whichever is lower in energy, when that energy is below $E_0$.  The difference between the d-wave $E_d$ and s-wave $E_s$ energies, when both are less than $E_0$ are shown in 
Fig.\ \ref{dsdiff}. These energy differences are usually smaller, sometimes much smaller, than $E_0-E_c$. 

     One awkward feature of Fig.\ \ref{cond1} is the fact that points of d- and s-wave condensate are interspersed, and often
next to one another in the $U-$density plane.  On the other hand, the energies of the optimal d-wave and s-wave states are
often so close together, differing by a fraction of one percent, that a change in lattice volume can reverse the energy ordering of the two states.  We have seen this in a number of points we have checked, whose d- and s-wave energies differ by O($10^{-4}$).  In such cases we find that a point of s-wave condensate on a $16\times 16$ lattice may convert, at the same density, to a d-wave condensate on a 
$24\times 24$ lattice, and vice versa.  Therefore only two statements can be made with any confidence:  condensates tend to lower the energy (compared to the Hartree-Fock ground state) at higher densities, generally at $\m>1, f>0.6$ with $U>4$, and spatial inhomogeneity is a necessary but not sufficient condition for their appearance.
 
\begin{figure}[h!]
 \center
 \includegraphics[scale=0.4]{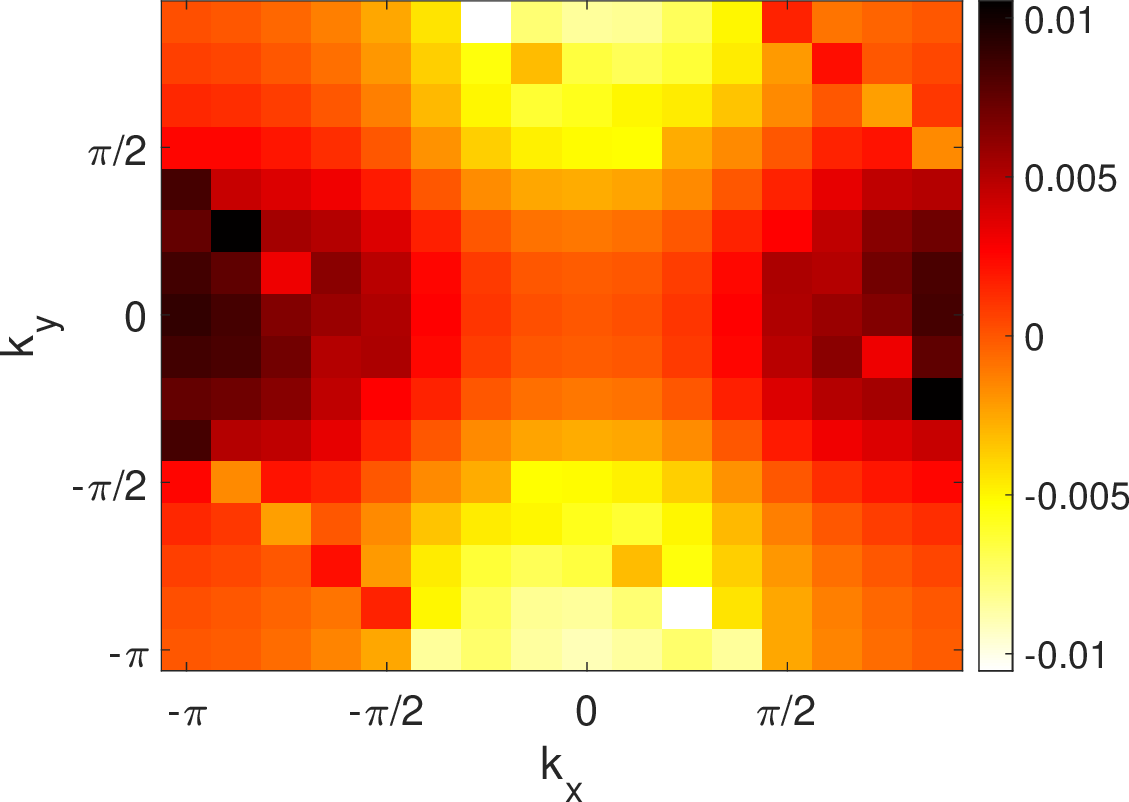}
 \caption{Momentum space condensate $P(k)$ at $U=8,\m=1.4,h=0.01$, where we find density $f=0.83$.}
 \label{pairU8}
 \end{figure}
     The momentum-space distribution of the condensate is quite different from the concentration at the Fermi surface seen in 
Fig.\ \ref{pairU2}. An example at  $U=8, \m=1.4, h=0.01$ and density $0.83$ which is displayed in Fig.\ \ref{pairU8}, is much more diffuse.  
In this case the $\E(h)=-1.802$, while $\E(0)=-1.789$.  The maximum amplitude of the condensate at small $U=2$ in 
Fig.\ \ref{pairU2}, which is concentrated at the Fermi surface, is much larger (by about a factor of forty) than the maximum amplitude of the condensate in the more diffuse result at $U=8$ shown in Fig.\ \ref{pairU8}.  But if one defines a measure of average condensate
magnitude
\beq
   C = {1\over L^2} \sum_k |P(k)| \ ,
\eeq
then the measures at $U=2$ and $U=8$ are $C=0.010$ and $C=0.0033$ respectively, differing only by a factor of three.  This is
simply because, although the maximum amplitudes at $U=8$ are small, the condensate is more diffuse in $k$-space as compared with the concentration at the Fermi surface at $U=2$, and there are small but significant contributions to $C$ over a greater momentum range.

\begin{figure}[h!]
 \center
 \includegraphics[scale=0.4]{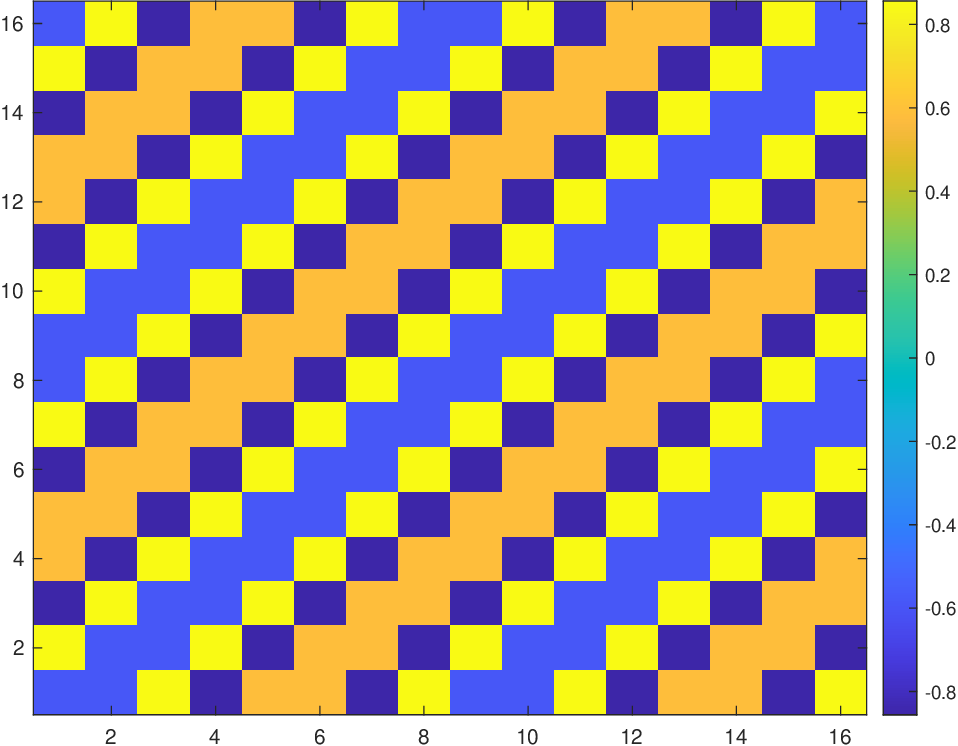}
 \caption{Spin average $D(x)$ on the $16\times 16$ lattice at parameters $U=8,\m=1.4,{h=0.01}$. $x$ and $y$ axes correspond to $x,y$ coordinates on the lattice, and the color of each square indicates the value of $D(x)$ at that position.}
 \label{geospin8}
 \end{figure}

The average spin at each site is
given by
 \bea
         D(x) &=& \langle c^\dg_\up(x) c_\up(x)\rangle  - \langle  c^\dg_\dn(x) c_\dn(x)\rangle \non \\
                 &=& \langle \psi_1^\dg(x) \psi_1(x) \rangle  + \langle \psi_2^\dg(x) \psi_2(x) \rangle - 1 \non \\
                 &=& \r(x,11) + \r(x,22) -1 \ ,
 \eea   
and the result for parameters $U=8, \m=1.4, h=0.01$ is shown in Fig.\ \ref{geospin8}.  It is clear in this case, as in all cases where we have found 
$\E(h)<\E(0)$ for some $h>0$, that the expectation value of spin densities is far from uniform.  Defining an average spin magnitude per site,
\beq
         m_{av} = {1\over L^2} \sum_x |D(x)| \ ,         
\eeq
we find that in all cases where  $\E(h)<\E(0)$ for some $h>0$, $m_{av}$ ranges from 0.30 to 0.94.  Averaging over all such cases, the value
is 0.63 with a standard deviation of 0.15 in this set.  Since $N_\up=N_\dn$, the spatial average of $D(x)$ must vanish.  Then the fact the spatial average $D(x)$ is zero yet the magnitude at each site deviates quite strongly from zero implies spatial inhomogeneity of some kind, as in Fig.\ \ref{geospin8}.  In contrast, running the numerical Hartree Fock program at the parameters $U=2, \mu=0.1$ as in Fig.\ \ref{pairU2}, where the $h=0$ state is lowest energy, we find $m_{av}=1.6 \times 10^{-16}$, i.e.\ near perfect spatial homogeneity.
 
 \section{Discusson and generalization}
 
     We have found that a slight variational improvement of the Hartree-Fock approach lowers the ground state energy, as compared to the unimproved state, at sufficiently large density and coupling,  and that these lower energy states are associated with either s- or d-wave condensates. Only spatially inhomogeneous states, with spin expectation values $D(x)$ significantly different from zero, seem to have this property.
For a review of the possible relation between stripes and superconductivity, see \cite{Tranquada_2020}. 
     
     A natural question is whether we can find a more systematic procedure for obtaining the optimal approximation to the ground state, rather than depending on the choice of a specific condensate-inducing term which, of course, may not be the ideal choice.  To improve matters, the general idea is still to combine the Hartree-Fock and variational approaches to optimize a single determinant approximation to the ground state.  A maximal (and impractical) procedure would be as follows:  On a finite $L\times L$ lattice at fixed $U$ and $\mu$, we apply the Hartree-Fock method to obtain a complete and orthogonal set of of one-particle wave functions 
$\{ \phi_n(x,\a), n=1,2,...,2L^2\}$.  The standard Hartree-Fock ground state is shown in \rf{Om}.  Any other Slater determinant can be obtained by replacing the set $\{\phi_n\}$ in \rf{Om} by a new set $\{\phi'_n = \U_{nm} \phi_m\}$, where
 $\U=\exp[i\g_i T_i]$ is a $2L^2 \times 2L^2$ unitary matrix which is a member of the $U(N)$ group, with $\{T_i\}$ the corresponding generators.  Then we may regard the
 $\g_i$ as variational parameters, and attempt to vary those parameters in such a way as to minimize the expectation value
 $\E = \langle \Om|H|\Om\rangle$, building $|\Om\rangle$, as before, out of a subset of the $\{\phi'_n\}$.  In principle,  if one could find the set of $\g$'s which minimizes $\E$, that would be the best approximation, by a single Slater determinant, to the true ground state of the Hubbard Hamiltonian.  In practice there will be a vast number of local minima, but aside from that this ``brute  force'' approach, with the number of variational parameters equal to
$2L^2$, seems extremely computation intensive.  But if it would be the case that the Hartree-Fock approach is only failing close to
the Fermi surface, then the problem may be more tractable.  For most one-particle states we let $\phi'_n=\phi_n$, and only choose a subset of $K$ one-particle states, in the immediate neighborhood of the Fermi surface, for which we consider $\phi' = \U \phi$.
This time, $\U$ is only a $K\times K$ matrix acting on states in the Hilbert space spanned by the $K$ Hartree-Fock one-particle states near the Fermi surface.
Of course this still leaves $K^2$ variational parameters, but one can imagine some stochastic or relaxation approach which would
converge to at least a local minimum.  For example, one could cycle through pairs of one particle states in the $K$ subset,
and apply at each stage an SU(2) transformation to obtain a new pair of states which are linear combinations of the original pair.  These could be accepted, as new one particle states in the Slater determinant, if they lower $\E$.  One would of course have to test the sensitivity of the final result to the choice of $K$, to ensure that enough states were chosen.

   We leave this last suggestion for future investigation.

\acknowledgments{We thank Richard Scalettar for helpful correspondence.  This research is supported by the U.S.\ Department of Energy under Grant No.\ DE-SC0013682.}  

\bibliography{hub}   
   
\end{document}